\documentstyle[epsf]{l-aa}
\begin{document}
\thesaurus{6(8.15.1; 8.22.1; 8.22.3; 11.09.1 IC 1613; 11.12.1; 11.19.5)}
\title{Variable stars in nearby galaxies.\thanks{Based on 
observations collected at ESO-La Silla}}
\subtitle{I. Search for Cepheids in Field A of IC 1613}
\author{E. Antonello, L. Mantegazza, D. Fugazza, M. Bossi, S. Covino}
\offprints{E. Antonello} 
\institute{Osservatorio Astronomico di Brera, Via E.~Bianchi 46,
       I--23807 Merate, Italy \\
~(elio,luciano,fugazza,bossi,covino@merate.mi.astro.it)}
\date{ Received date; accepted date }
\maketitle
\markboth{E. Antonello et al.: Field A of IC1613}
{E. Antonello et al.: Field A of IC1613}

\begin{abstract}
The first results are presented of a four-year program 
dedicated to the CCD observations of Cepheids in the nearby galaxy
IC 1613. The goal was to obtain good light curves for Fourier
decomposition and to detect shorter period Cepheids.
Since the program was carried out with a relatively small telescope,
the Dutch 0.9 m at ESO-La Silla, the observations were performed
without filter (white light), or $Wh$--band; the advantage of this technique 
is that the 
photon statistics correspond to that of $V$-band observations made with larger
telescopes than 2 m and similar exposure time. The effective 
wavelength of the $Wh$-band is intermediate between that of $V$ and $R$ 
bands for stars of A-G spectral type, for back-illuminated CCD detectors, 
therefore the photometric characteristics of variable stars (e.g. amplitudes) 
are generally analogous to those obtained with the standard technique.

Field A in IC 1613 has size 3{\farcm}8x3{\farcm}8. A total of 67 images
were obtained and the reduction was performed with DAOPHOT. More than
2900 stars were measured, and for about 1700 stars there are  
from 67 to 24 $Wh$ data points. Indications on the color of 739 bright 
stars were obtained also from $V$ and $R$ additional data.
The analysis revealed the presence of about 110 variable stars. 
The detected population I Cepheids are 43; 9 Cepheids were already known 
from previous works, while most of the new
stars have a short period $P$. We remark the following results: 
a) for stars with $P \ga 5$ d and sufficient phase coverage it is 
possible to perform good Fourier decomposition with resulting standard 
deviation of the fit of  0.02 - 0.04 mag; b) there
are several Cepheids with relatively small amplitude, and most of them are  
(probable) first overtone mode pulsators; c) the faintest detected Cepheids 
have m$_V \sim 23$.  No double-mode Cepheid has been 
found, probably because the precision and sampling of the data are not 
sufficient for the detection. Furthermore, at least 5 population II Cepheids
and at least 8 eclipsing binaries have been observed. The other variable stars 
are probable long period, semiregular and irregular variables.

A comparison with results of other massive CCD photometric projects
dedicated to the detection of variable stars shows some advantages of the 
observations in white light for fully exploiting the capabilities of
relatively small telescopes. A suggestion is made on how to use these results
for distance determinations.

\end{abstract}


\begin{figure*}
\epsfysize=15truecm
\epsffile[10 150 420 660]{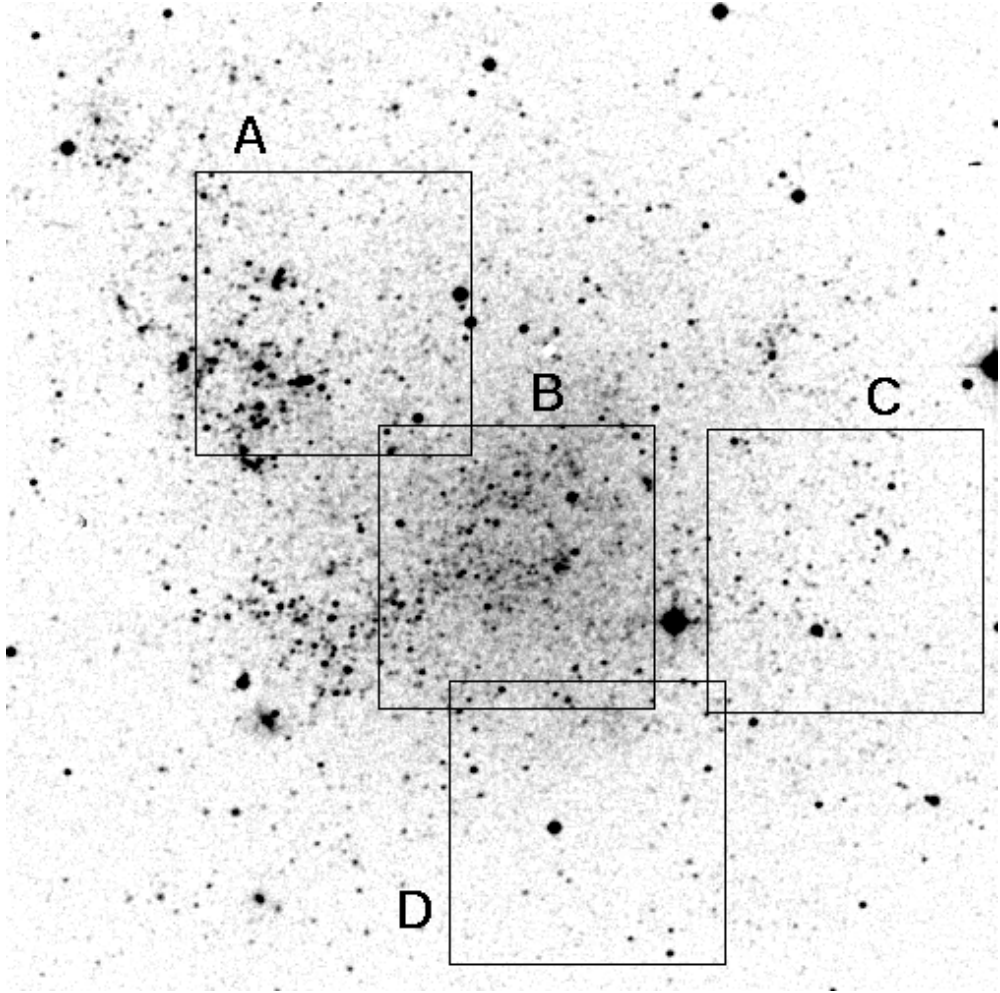}
\caption[ ]{The four surveyed field of IC 1613. The background image comes
from the STScI Digitized Sky Survey (north is up and east to the left), and
measures approximately 13{\farcm}5 on each side
}  
\end{figure*}

\begin{table*}
\caption[]{Log of observations of Field A of IC 1613}
\begin{flushleft}
\begin{tabular}{|llllll|llllll|}
\hline\noalign{\smallskip}
 &date&Hel.J.D.&Airmass& FWHM & N & & date & Hel.J.D.  &Airmass& FWHM & N \\
  &  & 2450000.+ &  &   ('') &   & &  & 2450000.+ &  &  ('') &  \\
\hline\noalign{\smallskip}
 1 &1995 oct 20  &   11.69 & 1.18 & 1.5 & 2162 &
34 &1997 oct 27  &  749.67 & 1.19 & 1.6 & 1685\\
 2 &1995 oct 21  &   12.75 & 1.37 & 1.7 & 1372 &
35 &1997 oct 28  &  750.61 & 1.19 & 1.5 & 1806\\
 3 &1995 oct 22  &   13.74 & 1.35 & 1.5 & 1709 &  
36 &             &  750.71 & 1.36 & 1.8 & 1589\\
 4 &1995 oct 23  &   14.59 & 1.27 & 1.6 & 1930 &
37 &1997 oct 29  &  751.60 & 1.20 & 1.8 & 1252 \\
 5 &1995 oct 25  &   16.62 & 1.19 & 1.4 & 2105 &
38 &             &  751.70 & 1.29 & 2.0 & 1324 \\
 6 &1995 oct 26  &   17.62 & 1.19 & 1.7 & 1870 &
39 &1997 oct 30  &  752.61 & 1.19 & 1.6 & 1941\\
 7 &1996 aug 06  &  302.87 & 1.18 & 1.8 & 1217 &
40 &             &  752.70 & 1.29 & 1.8 & 1340\\
$V$&             &  302.89 & 1.18 & 1.6 &  687 &            
41 &1997 oct 31  &  753.60 & 1.19 & 1.7 & 1303\\
$R$&             &  302.91 & 1.22 & 1.6 &  827 &
42 &             &  753.63 & 1.18 & 1.4 & 1814\\
 8 &1996 aug 08  &  304.79 & 1.25 & 1.9 & 1179 &
43 &             &  753.72 & 1.40 & 1.6 & 1724\\
 9 &             &  304.80 & 1.24 & 1.4 & 1930 &
44 &1997 nov 01  &  754.61 & 1.18 & 2.5 & 829\\
10 &1996 aug 09  &  305.80 & 1.23 & 1.5 & 1214 &
45 &             &  754.71 & 1.33 & 2.1 & 1127\\
11 &1996 aug 10  &  306.82 & 1.18 & 1.6 & 1428 &
46 &1998 jul 25  & 1020.90 & 1.17 & 1.7 & 1593\\
12 &1996 aug 11  &  307.77 & 1.29 & 2.4 &  748 &
47 &1998 jul 26  & 1021.89 & 1.17 & 1.5 & 1451\\
13 &             &  307.91 & 1.30 & 2.0 & 1465 &
48 &1998 jul 30  & 1025.89 & 1.17 & 2.1 & 1139\\
14 &1996 aug 12  &  308.86 & 1.20 & 1.9 & 1246 &
49 &1998 jul 31  & 1026.89 & 1.17 & 1.5 & 1812\\
15 &1996 aug 13  &  309.83 & 1.17 & 2.2 & 1082 &
50 &1998 sep 30  & 1087.71 & 1.17 & 1.4 & 1638\\
16 &1996 oct 14  &  370.57 & 1.52 & 1.5 & 1247 &
51 &1998 oct 01  & 1088.74 & 1.19 & 1.6 & 1781\\
17 &             &  370.71 & 1.19 & 1.6 & 1355 &
52 &1998 oct 02  & 1089.73 & 1.19 & 1.6 & 1487\\
18 &1996 oct 15  &  371.59 & 1.39 & 1.6 & 1469 &
53 &1998 oct 08  & 1095.67 & 1.17 & 1.6 & 822\\
19 &             &  371.72 & 1.27 & 1.3 & 1784 &
54 &1998 oct 09  & 1096.71 & 1.18 & 1.6 & 1535\\
20 &1996 oct 16  &  372.70 & 1.18 & 1.3 & 2068 &
55 &1998 oct 10  & 1097.67 & 1.18 & 1.5 & 1564\\
21 &1996 oct 17  &  373.61 & 1.26 & 1.3 & 1826 &
56 &             & 1097.77 & 1.33 & 1.6 & 1948\\
22 &             &  373.75 & 1.34 & 1.4 & 1001 &
57 &1998 oct 11  & 1098.64 & 1.21 & 1.7 & 1718 \\
23 &1996 oct 18  &  374.57 & 1.49 & 1.5 & 1444 &
58 &             & 1098.74 & 1.25 & 1.5 & 1480 \\
24 &             &  374.69 & 1.18 & 1.2 & 2307 &
59 &1998 oct 13  & 1100.67 & 1.17 & 1.5 & 1453 \\
25 &1996 oct 19  &  375.61 & 1.25 & 1.2 & 1461 &
60 &             & 1100.77 & 1.37 & 1.8 & 1547 \\
26 &             &  375.75 & 1.33 & 1.2 & 2753 &
61 &1998 oct 14  & 1101.74 & 1.27 & 2.2 & 1458 \\
27 &1996 oct 20  &  376.62 & 1.23 & 1.6 & 1496 &
62 &1998 oct 15  & 1102.59 & 1.34 & 1.6 & 1511\\
28 &             &  376.75 & 1.35 & 1.7 & 1473 &
63 &             & 1102.69 & 1.18 & 1.6 & 1949\\
29 &1996 oct 21  &  377.61 & 1.25 & 1.6 & 1195 &
64 &1998 oct 16  & 1103.59 & 1.34 & 1.4 & 1776\\
30 &             &  377.74 & 1.31 & 1.7 & 1337 &
65 &             & 1103.69 & 1.18 & 1.5 & 1400\\
31 &1996 oct 22  &  378.60 & 1.26 & 1.5 & 1398 &
66 &1998 oct 17  & 1104.66 & 1.17 & 1.3 & 2066\\
32 &             &  378.68 & 1.18 & 1.4 & 1400 &
67 &1998 oct 18  & 1105.67 & 1.17 & 1.3 & 1921\\
33 &1997 oct 27  &  749.54 & 1.46 & 2.0 & 654 &
   &             &         &      &     &     \\
\noalign{\smallskip}
\hline
\end{tabular}
\end{flushleft}
\end{table*}

\section{Introduction}
\subsection{Motivation}
Cepheids are variable stars which are used to measure distances of
galaxies in the Local Group and nearby clusters (e.g. Madore et al. \cite{m1}), 
and are the primary calibrator for the secondary standard candles that are applied 
at much greater distances (e.g. Jacoby et al. \cite{ja}). However, 
they are not only fundamental stars as primary distance indicators, but 
are also an essential tool for testing the theories on the internal 
constitution of stars and stellar evolution. The importance of double--mode 
Cepheids for the revision of stellar opacities is well known: after the 
suggestion of Simon (\cite{s82}), the OPAL (e.g. Iglesias, Rogers \& Wilson 
\cite{irw}) and OP (e.g. Seaton et al. \cite{sea}) projects produced new 
opacities which allowed to solve the long-standing problem of double-mode 
Cepheid period ratios (e.g. Moskalik et al. \cite{mbm}). These opacities were 
then generally adopted by theorists working with stellar evolution codes. 

There are several problems yet to be solved. The radiative codes used for
constructing pulsation models proved to be incapable of agreement with 
observations when applied to the comparison of Cepheid characteristics in
Galaxy and in Magellanic Clouds (e.g. Buchler \cite{buc}). 
The fact that resonances among the pulsation modes give rise to observable
effects on the light curves can be exploited to put constraints on the pulsational 
models and on the mass-luminosity relations. The best known of these resonances 
occurs in the fundamental Cepheids between the fundamental and the second overtone 
mode ($P_0/P_2=2$) in the vicinity of a period $P_0 \sim 10$ d and it is at the 
origin of the well known Hertzsprung progression of the bump Cepheids
(e.g. Simon and Lee \cite{sl}). In the first overtone mode Cepheids another
resonance occurs between the first and the fourth pulsation modes
($P_1/P_4=2$; e.g. Antonello \& Poretti \cite{ap}; Antonello, Poretti and 
Reduzzi \cite{apr}). When these resonances observed in Cepheids of Galaxy and
Magellanic Clouds are used to constrain purely radiative models, one obtains 
stellar masses that are too small to be in agreement with stellar evolution 
calculations. According to Buchler et al. (\cite{bu2}), it has become clear 
that some form of convective transport and of turbulent dissipation is needed 
to make progress.

The study of Cepheids in nearby galaxies is of fundamental importance for 
understanding the effects of different metallicity and corresponding
mass--luminosity relations on the pulsational characteristics through 
the detection of structures in the {\it Fourier parameter - 
period} diagrams of fundamental, first overtone and possibly double--mode 
(Poretti \& Pardo \cite{pp}) and second overtone mode Cepheids 
(Antonello \& Kanbur \cite{ak}; Alcock et al. \cite{macho2}), and their comparison 
with the galactic Cepheids and the model predictions. The CCD differential 
photometric precision allows to get accurate Fourier parameters of Cepheid 
light curves, and also to discover several new Cepheids with small amplitude.
Massive CCD photometry of nearby galaxies such as NGC 6822 and IC 1613 
was attempted several years ago by E. Schmidt and collaborators 
(Schmidt \& Spear \cite{sch}), but apart from a preliminary report, no 
complete study was published. The MACHO, EROS and OGLE projects dedicated to the 
detection of microlensing events in the direction of Magellanic Clouds produced 
enormous amount of data on variable stars in these galaxies (e.g. Welch et al. 
\cite{we}; Beaulieu \& Sasselov \cite{bs}; Udalski et al. \cite{uda}). 
More recently, the project DIRECT was 
dedicated to the massive CCD photometry of M31 (and M33) with the purpose of 
detecting Cepheid and eclipsing binaries for direct distance determination of 
these galaxies (e.g. Kaluzny et al. \cite{kal}).

The purpose of our project was to obtain good light curves of Cepheids for
extending the comparison of the characteristics of these stars
in different galaxies. In order to exploit the telescope time and reach the 
faintest luminosities, our strategy was to observe in white light, i.e.
without filter; the results confirm that in this way the differential photometry 
precision for the Ducth 0.91 cm telescope at ESO--La Silla is roughly comparable 
with that obtained with 2 m-class telescopes, Johnson V-filter and similar
exposure times. 

In the present work we discuss observations and reduction methods, we
present the first results concerning population I Cepheids and other variable
stars. Subsequent papers will be dedicated to the analysis of population I and 
II Cepheids, long period and irregular variables and eclipsing binaries.

\subsection{IC 1613}
The irregular galaxy IC 1613 [$\alpha=1^h 02^m 16^s$ (1950), $\delta= 
+1\degr 52'$ (1950), l=130\degr, b=--61\degr], was studied by Baade, 
but his extensive results were never published. 
Baade found 59 variables in plates taken with the Mount Wilson
60 inch and 100 inch reflectors between 1929 and 1937. Light curves for 24 
of the confirmed Cepheids had been completed by him before his death in 1960.
These data, reduced to a new photometric scale, were published by Sandage 
(\cite{san}), who discussed the apparently anomalous slope of the $PL$
relation. The cause of this shallower slope than that of Cepheids in other 
galaxies was interpreted differently by various authors; the reasons for such 
interest was that if the slope was significantly flatter
for IC 1613, then the assumption that a universal $PL$ relation exists 
was seriously called into question. Freedman (\cite{fre1}) discussed the 
case with new CCD $BVRI$ data, and noted that, for data fainter than 21 mag,
the photographic photometry was significantly brighter than CCD data.
This divergence contributed to the difference in the appearance of the $PL$
relation.  Sandage (\cite{sa2}) discussed another cause, that is the
stochastic effect of small sample statistics for the few longest-period
variables. An additional 16 Cepheids were subsequently presented by Carlson 
\& Sandage (\cite{cs}), and the authors remarked the possible large number of
short period Cepheids. The conclusion of these studies is that there are
no differences in the slope of the $PL$ relation of Cepheids in IC 1613
with respect to that of other galaxies.
 
From $BVRI$ observations (Freedman, \cite{fre1}), Madore \& Freedman (\cite{mf}) 
derived a total mean reddening of $E$($B$--$V$)=0.02 mag, and a true distance 
modulus of 24.42${\pm}$0.13 mag, corresponding to a distance 
of 765 kpc. Madore \& Freedman  (\cite{mf}) suggest that the best place
for work on intrinsic calibration problem of the Cepheid distance scale is
not the Magellanic Cloud system but IC 1613, because the foreground reddening 
to this galaxy is very low and probably quite uniform, the extinction 
internal to IC 1613 appears to be quite small and the crowding of stellar 
images are relatively low. Freedman (\cite{fre1}) mentions other points of 
interest of this galaxy, and the need of better data on its Cepheids. 
IC 1613 has very low metallicity, less than SMC, and both are 
important galaxies for calibrating the $PL$ relation, but the latter has a 
complicated extended geometry; therefore the former could be even more 
important in this regard than was once thought. With more and better 
data, the Cepheids of IC 1613 could provide the low-metallicity anchor point
for a calibration of the $PL$ relation.

\begin{table}
\caption[]{Comparison of effective wavelengths (nm)}
\begin{flushleft}
\begin{tabular}{llll}
\hline\noalign{\smallskip}
Spectral type & $V$ & $R$ & $Wh$ \\
\noalign{\smallskip}\hline\noalign{\smallskip}
B & 543 & 659 & 503 \\
A & 545 & 667 & 548 \\
F & 547 & 671 & 576 \\
G & 548 & 676 & 601 \\
\noalign{\smallskip}
\hline
\end{tabular}
\end{flushleft}
\end{table}

\section{Observations}

The observations were performed with the direct CCD camera attached to the 
Dutch 0.91m telescope of the La Silla Astronomical Observatory (ESO) 
during six runs from October 1995 to October 1998. The available CCD 
detector was the ESO chip No. 33, which is a TEK CCD with 512x512
pixels, pixel size of $27{\mu}m$ and spatial resolution of 0{\farcs}44,
providing a field of view of 3{\farcm}77x3{\farcm}77. Given the limited size 
of the field of view, the need to observe not too far from the meridian and at 
the same time to be able to get two images of the same field in the same night,
we were forced to limit our programme to 4 selected fields of IC1613.
The fields are displayed in Fig. 1. Most of the observations were performed 
without filter (white light, hereinafter $Wh$) in order to get the best photon
statistics for the study of faint Cepheid light curves; few images were 
taken in Johnson $V$ and $R$ filter for comparison purposes with other works 
and to get an indication on the colors of the relatively bright stars.
Bias and twilight flat field frames were gathered in each useful night. 

In this paper we present the results regarding Field A, which is an
interesting field because it contains the largest number of previously known variable 
stars. During the allocated telescope time the field has been observed 
for 46 nights and a total of 67 images were collected. The complete
log of the observations is reported in Table 1. The table contains, for each image,
the date of the beginning of the night, the Heliocentric Julian date of midexposure,
the mean FWHM PSF value (which takes into account both seeing and possible 
non--perfect focussing of the telescope), the airmass, and the number of 
stars detected in the image, which is an indication of its overall goodness
because it depends both on the PSF (point-spread-function) and sky background 
level. Each $Wh$, $V$ and $R$ image is the sum of two or three successive 
exposures for a total of 1800 sec; only for the image No. 61 the total exposure 
was shorter (600 sec).

The observations without filter imply the dependence of the resulting photometry 
on: 1) the adopted instrumentation; b) a color term which is related to the 
different airmasses. The effective wavelengths of $V$ and $R$ bands are compared 
in Table 2 with that of $Wh$ band, for stars with B, A, F and G spectral type
observed with the adopted instrumentation, that is optical telescope and 
back-illuminated CCD-TEK detector. The $Wh$--band effective wavelength for 
late-type stars is intermediate between that of $V$ and $R$ bands. The 
color-effect due to different airmasses in the present case is negligible
(in comparison with the expected photometric precision) since the observations 
were performed not far from meridian.

\section{Data Reduction}
Bias subtraction and flat field corrections were performed using IRAF package
(Tody 1993). It was necessary to correct the images for the slightly different 
pointing of the telescope and orientation of the CCD camera in the different runs. 
Therefore they were shifted and rotated with the ESO/MIDAS command 
REBIN/ROTATE with respect to the image No. 26, which is one of
the best images and which was used as template. For this reason, the stars 
near the borderline have usually less measurements than those in the central 
part of the field, and the field actually surveyed is sligthly larger
than the nominal one, that is about 3{\farcm}84x3{\farcm}84.

\subsection{Photometry}
The stellar photometry was performed by means of the IRAF/DAOPHOT package
(Stetson 1987; Davis 1994). For each image a prelimary list of objects was detected with 
DAOFIND, and a prelimary aperture photometry was performed with DAOPHOT. 
In order to evaluate the point-spread-function, a group of stars was selected 
with PSTSELECT and then checked visually one by one. 
The point-spread-function model was then iteratively computed with the PSF--command 
using about 20 stars for each image. Due to the smallness of the field, a constant 
PSF model consisting of a gaussian plus a single empirical look--up table was 
adopted. Finally the photometry of all the selected stars was derived by means 
of ALLSTAR. New stars were then searched in the residual image, added to the 
previous list, and then ALLSTAR was executed again on this list.
This procedure was finally repeated once again. The residual image that we 
got after the third analysis with ALLSTAR was generally clean, with no 
evident stellar images; only some residuals near the loci of the brightest 
stars, HII regions and galaxies were present. The tables containing the lists 
of the detected stars in each image were cross--correlated in order to look 
for the objects in common and reject the spurious ones. Two objects in two 
different images were considered to be the same star if the separation of 
their centers was less than 1 pixel.

The same reduction procedure was adopted for both $Wh$ and $V,R$ images.

\begin{figure}
\epsfysize=11truecm
\epsffile[30 340 380 690]{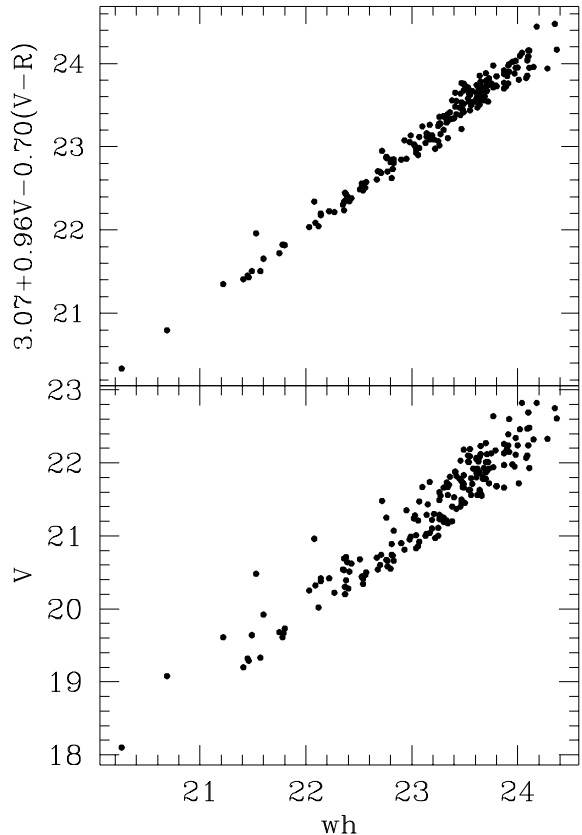}
\caption[ ]{Correlations among the photometric parameters of 195
selected stars. Lower panel: $V$ against instrumental $wh$; upper panel:
best fit given by eq. 2
}  
\end{figure}

\subsection{Calibration}
DAOPHOT produced a set of {\em instrumental wh} magnitudes 
for the stars in each frame, which could not be reduced to a standard system.
The procedure for deriving a homogeneous magnitude scale was an
iterative one, based on the 104 stars detected in all the frames.
Let $wh_{ij}$ be the magnitude value of the star $i$ 
in the frame $j$, $<wh_j>=\sum_i wh_{ij}/n$ the mean value for the
frame $j$, $n$ the number of stars, $<wh>=\sum_j<wh_j>/k$
the global mean value and $k$ the number of frames. At the first step, 
the mean values are computed and then a new $wh_{ij}^0$ value is calculated 
with the formula 
\begin{equation}
wh_{ij}^0=wh_{ij}-<wh_j>+<wh>.
\end{equation}
At each subsequent step the mean values of the time series 
$<wh_i>=\sum_j wh_{ij}^0/k$ are calculated using the new
$wh_{ij}^0$ values, excluding from each time 
series $i$ the most deviating point 
if the deviation is larger than 2.5$\sigma$ from $<wh_i>$, 
and replacing this 
point with the corresponding $<wh_i>$; then the new mean values
$<wh_j>$ and $<wh>$ are recalculated. The final result of this iteration 
is the correcting term, $-<wh_j>+<wh>$, applied to each of the 
original data points. The number of steps is fourteen, however for most 
of the stars very few iterations are sufficient for obtaining stable mean values.

\begin{figure}
\epsfysize=8.5truecm
\epsffile[10 150 380 720]{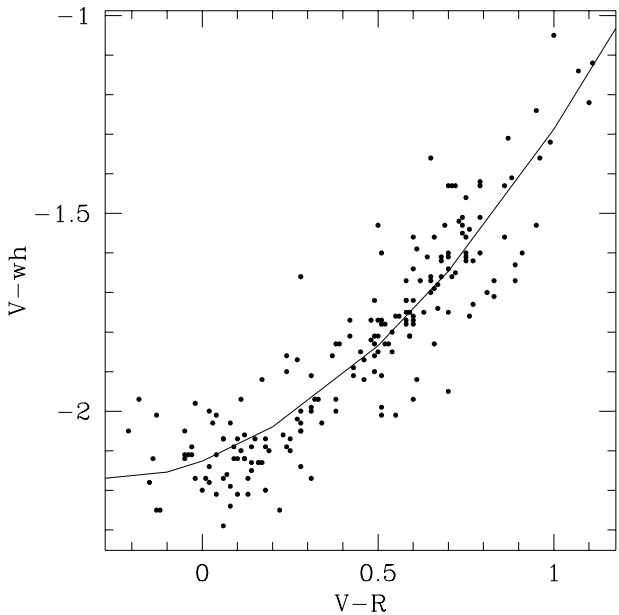}
\caption[ ]{$V$--$wh$ against $V$--$R$ for 195 selected stars 
}  
\end{figure}

\begin{figure}
\epsfxsize=14truecm
\epsffile[50 350 450 690]{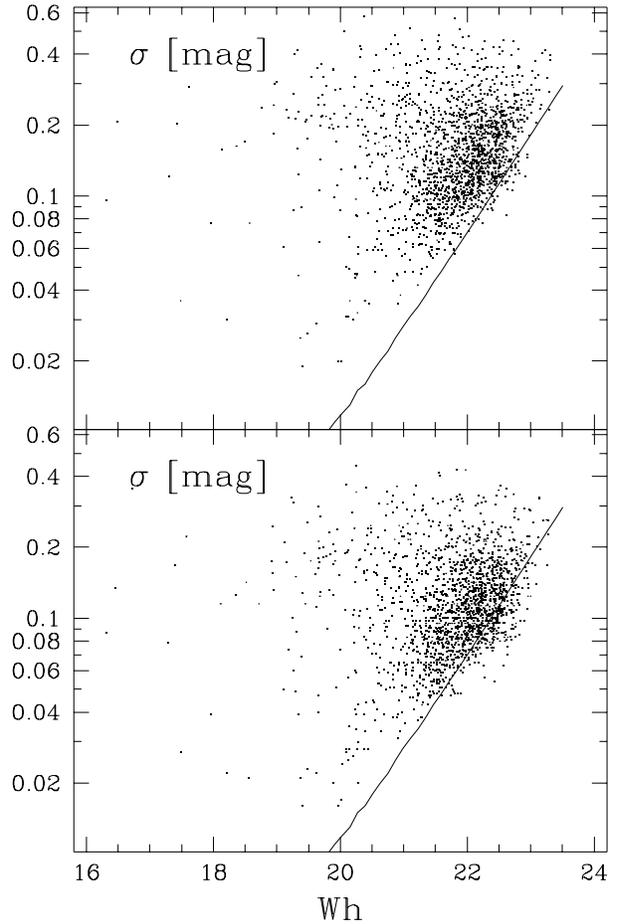}
\caption[ ]{Standard deviation $\sigma$ of the $Wh$ observations (logarithmic
scale) for stars with at least 24 data points and $\sigma<0.6$. 
Upper panel: all the observed points have been included, some of which are 
largely deviating, owing e.g. to crowding problems; lower panel: the four most 
deviating points have been excluded from the time series of each star. 
The continuous line is a rough theoretical estimate of the expected external error.
}
\end{figure}

Since Field A partially overlaps one of the fields observed by
Freedman (\cite{fre2}), we used the 158 common stars to tie our $V$ and $R$ 
observations to the standard $VR$ system. We got the color
index $V$--$R$ for 512 stars from our data; for an additional
227 stars, which were not detected in both $V$ and $R$ frames but were
detected in $wh$ frames, we adopted the $V$--$R$ value given by 
Freedman (\cite{fre2}).

We selected a sample of stars for constructing a $V, R, Wh$ 
system, useful for the discussion of the photometric results. The calibrating
stars were selected according to the following criteria: a) more than 62
observed $wh$ data points  per star, b) stellar nonvariability or 
low scatter of data points, c) known $V$ and $R$ data; criteria a) and b) 
were needed in order to get a sample free of problems related 
to crowding. The resulting number of stars was 195, which gave the following 
statistical relations between $wh$, $V$ and $V$--$R$ 
\begin{equation}
wh=3.07+0.96V-0.70(V-R)
\end{equation}

\begin{equation}
V-wh=-2.13+0.33(V-R)+0.51(V-R)^2,
\end{equation}
with rms residuals of 0.11 and 0.10 mag, respectively.
Fig. 2 shows $V$ (lower panel) and estimated $wh$ from eq. 1 (upper panel)
against observed $wh$, and Fig. 3 shows $V$--$wh$ against $V$--$R$.
We derived the zero-point $a_0$ of the final $Wh$ magnitude scale,
$Wh=wh+a_0$ from eq. 2 assuming that $V$--$Wh=0$ when $V$--$R=0$, that is
$a_0=-2.13$ or
\begin{equation}
Wh=wh-2.13.
\end{equation}
The nonlinearity of eq. 3 depends on the large $Wh$-bandwidth;
$V$--$wh$ appear to be more sensitive to $T_e$ than $V$--$R$ for cooler stars,
and less sensitive for hotter stars.
We will use occasionally the colour index $V$--$Wh$ when the $R$ measurement 
will not be available; we just note that, as a first approximation, from the
linear correlation between $V$--$Wh$ and $V$--$R$ we have
$V-Wh\sim0.6(V-R)$. Analogously, $Wh-R\sim0.4(V-R)$.

In Fig. 4 we have reported the external error (or standard deviation)
$\sigma$ for about 1700 stars with at least 24 data points (and $\sigma<0.6$), 
against the mean value $Wh$ of the star; in the upper panel, the standard
deviation was calculated including all the data points, some of which are 
rather scattered; in the lower panel, the four most deviating points of each 
star have been excluded. The continuous line is a rough
theoretical estimate of the expected external error.

\begin{figure}
\epsfysize=9truecm
\epsffile[10 150 380 720]{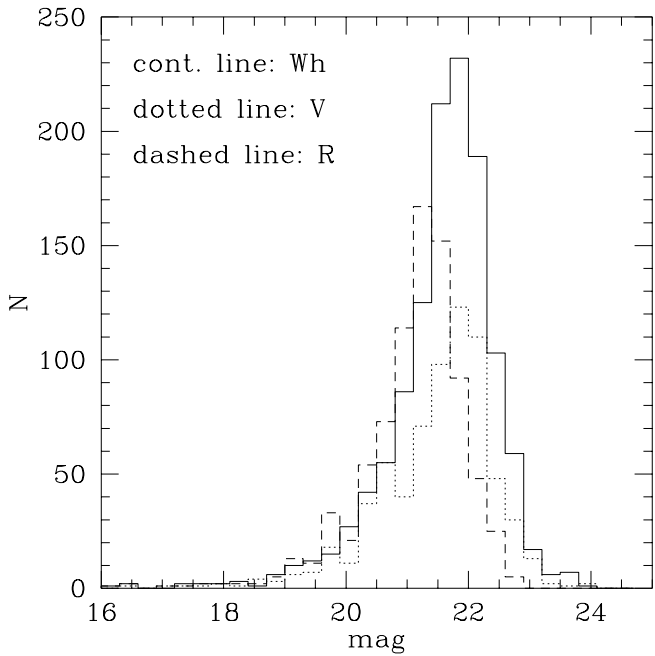}
\caption[ ]{Distribution in $Wh$, $V$ and $R$ of stars in Field A
for images taken with similar observing conditions
}  
\end{figure}

\section{Data Analysis}

\subsection{Star numbers}

It is interesting to compare the number of detected stars in $Wh$ band
with those in $V$ and $R$ bands. The total number of stars with at least one 
$Wh$ measurement is 2927; however, since the field was not exactly the same
during all the nights, the detected stars are located in a region of IC1613
which is actually slightly larger than 3{\farcm}77x3{\farcm}77. 
Moreover, the number of detections depend on the seeing conditions 
and sky background (see Table 1). We have 
considered therefore the $Wh$ frame taken in the same night of $V$ and 
$R$ frames, with similar seeing conditions. The resulting number of
detections is 1217 for $Wh$, 687 for $V$ and 827 for $R$. Fig. 5 shows
the corresponding distribution. 
\begin{figure}
\epsfysize=8.5truecm
\epsffile[10 150 380 720]{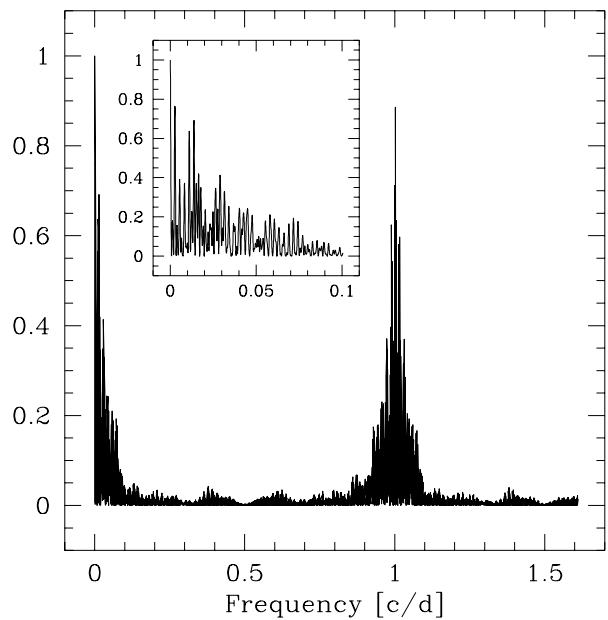}
\caption[ ]{Spectral window of 1995-1998 $Wh$ data of IC 1613. The inset
shows the fine structure of the main peak
}  
\end{figure}

\begin{figure}
\epsfxsize=17truecm
\epsffile[60 350 500 690]{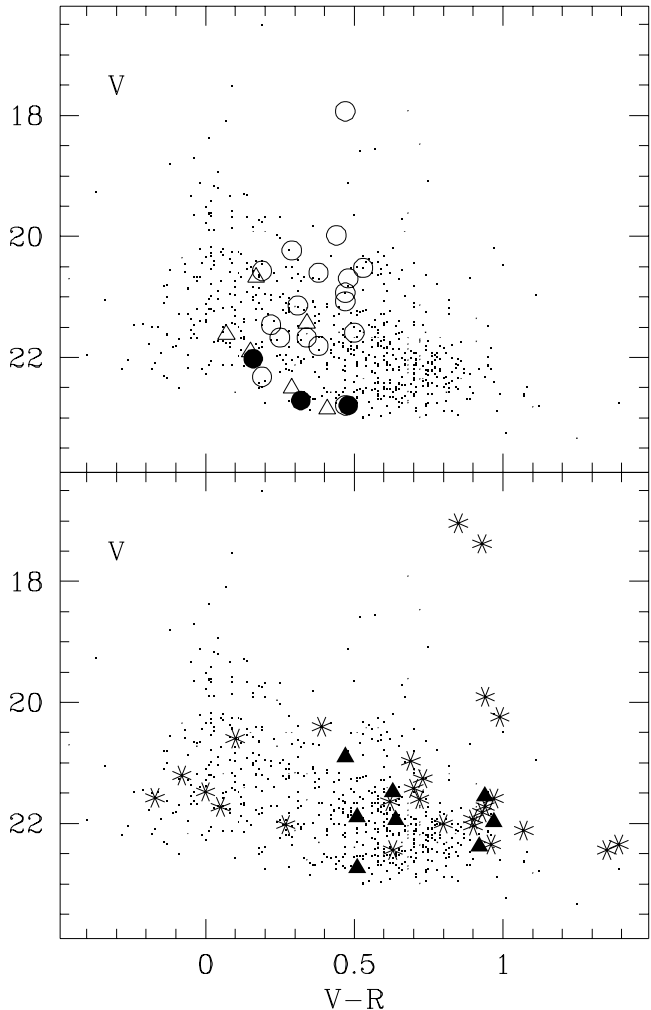}
\caption[ ]{Color-magnitude ($V$--$R$,$V$) diagram. Upper panel: 
{\em open circles:} population I Cepheids of fundamental mode; 
{\em filled circles:} population I Cepheids of first overtone mode;
{\em open triangles:} eclipsing binaries. 
Lower panel: {\em filled triangles:} periodic variables (those with
$V$--$R$ less than about 0.6 are population II Cepheids);
{\em asteriscs:} other semiregular and irregular variables
}  
\end{figure}

\subsection{Variable stars}
Different criteria for the detection of variability were
adopted for comparing the capabilities of the various methods, particularly 
in the difficult cases given by the uncertainties due to crowding problems.

Firstly we used the variability index $J$ (Stetson \cite{ste}). 
For each star the pairs of observations were considered, each with a weight 
$w_k$, where $k$ indicates the pair of observations $i_k,j_k$. If the time 
separation between two subsequent observations was less than about three hours 
hour, they were considered as a pair. When $i_k \neq j_k$ the weight was 
$w_{k}$=1.0, while when $i_k = j_k$, $w_{k} = 0.25$. In this way, longer 
sequences of closely spaced observations had larger weight than sequences 
with similar number of observations but largely separated in time.
The index $J$ was redefined in order to take into account how many times a 
given star was measured, $J_S = J \sum W / W_{max} $, where $W_{max}$ 
is the total weight a star would have if measured in all the images
(see Kaluzny et al. \cite{ka1}). As expected (Stetson \cite{ste}), 
most of the stars have $J_{S}$ values which are close to zero. 
The adopted threshold, $J_{S,min} = 0.5$ allowed the selection of 
136 candidate variable stars.

For another test for variability we considered 1491 light curves consisting of 
not less than 34 points, i.e. corresponding to stars which can be identified in the 
majority of the frames. They were checked for variability by means of 
two different methods, that is, their variances have been compared with 
two different noise estimates. At first, the noise component of a light 
curve was defined as the least variance found among the fainter objects than 
the examined one: a star was regarded as variable if its light variance, 
computed without taking the 10\% most scattered measurements into account, 
exceeded 10 times this level. The white noise component of
each time series was evaluated also from the root--mean--square
difference between closely consecutive data, i.e., in our case, between
measurements performed during the same night. When the light variance was
inconsistent (i.e. larger than 3\( \sigma \)) with this noise definition, 
the object was classified as a variable star. These approaches have to be 
considered as complementary. The first one is based on no more than a rough 
estimate of the noise, which doesn't depend only on the measured magnitude 
but also, e.g., on the crowding in the image. On the other hand, rapid 
variations with time scales of some hours may escape detection with the 
second method. Combining both approaches, more than 250 candidate 
variable stars were singled out for a further detailed analysis.

Finally, some time was also spent for analyzing the data set on a 
star-by-star basis. A simple program was developed which computed the
variance reduction for the time series, identified the maximum peak 
in the power spectrum, and showed the changes of peak and variance reduction
when taking off progressively the most deviating points from the time series. 
A good indicator of variability was the stability of the power spectrum 
peak, even when the variance reduction was not very significant. 
In this way it was possible to detect variable stars with 
relatively small amplitude.

From the comparison of the three approaches we got the indication that,
for an uneven data sampling as in the present case, the automatic methods
should adopt very low threshold levels in order to detect variable stars
which have low amplitude/noise ratio: on the one hand, this low threshold 
level yield also a large number of candidates which turn out to be nonvariable 
stars, and on the other hand some variable stars were found with the star-by-star
analysis below such thresholds.
\begin{figure*}
\epsfxsize=14truecm
\epsffile[30 180 450 690]{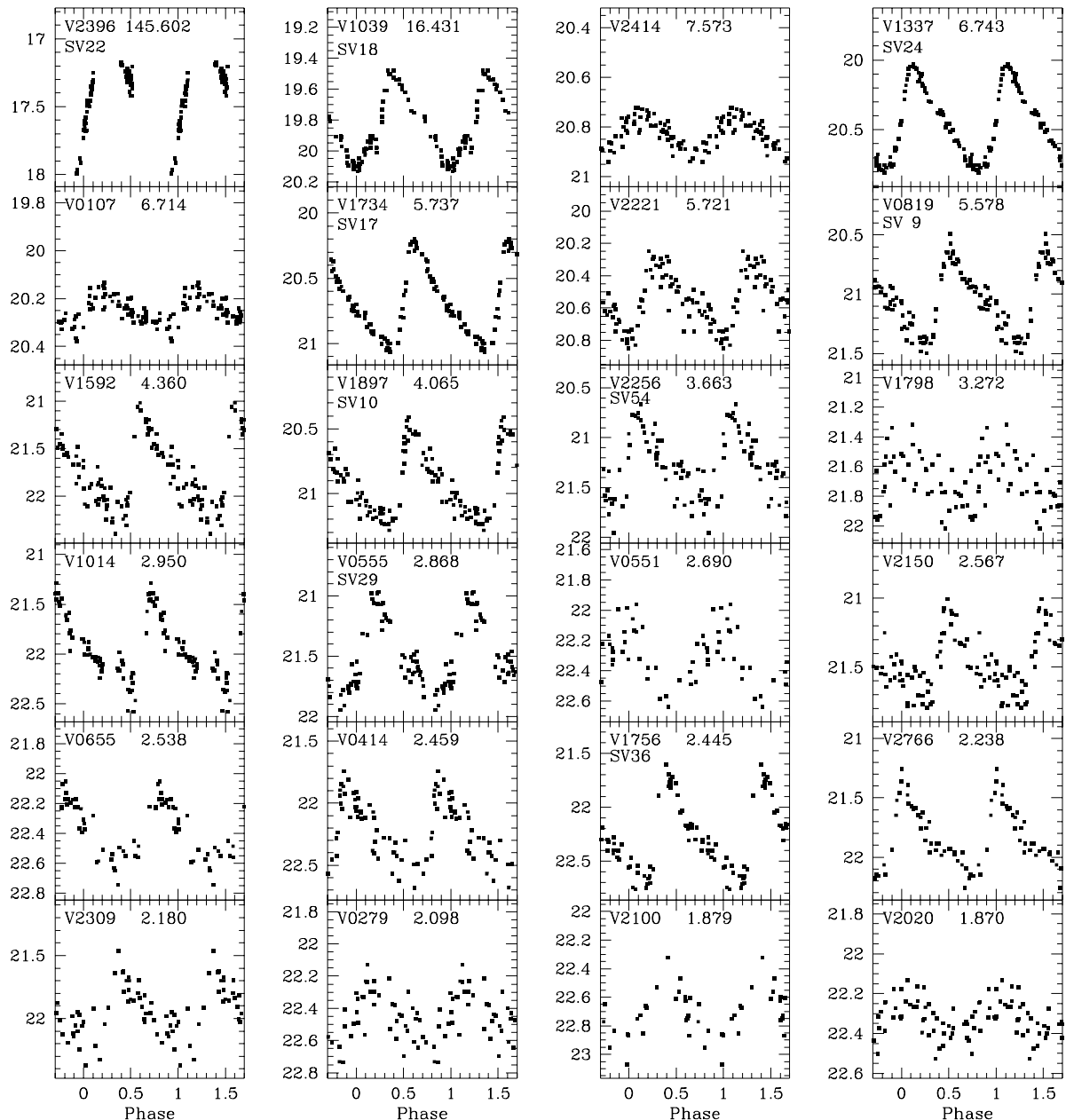}
\caption[ ]{Cepheid $Wh$ light curves. For each star, the identification
number and the period are reported. Note the different magnitude
scales. The $V$--$Wh$ range of Cepheids is 0.1 - 0.4 mag, and therefore
for these stars $V$ is larger than $Wh$ by some dex}  
\end{figure*}

\begin{figure*}
\epsfxsize=14truecm
\epsffile[30 250 450 690]{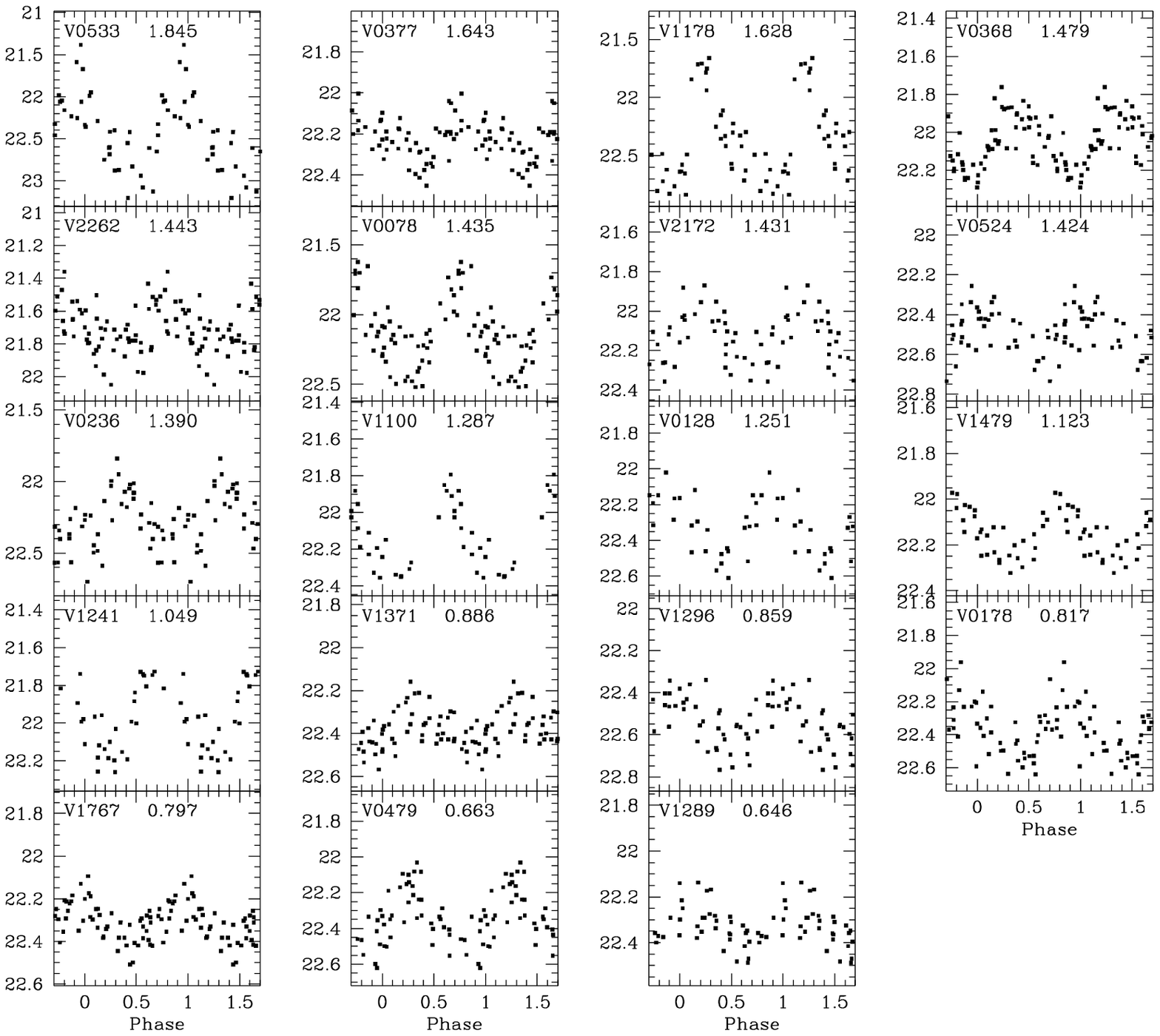}
\caption[ ]{Cepheid $Wh$ light curves ({\em contd.})
}  
\end{figure*}

The uncertainties in the analysis for the detection of variable stars
and the determination of their periods are mainly related to the number
of bad points (characterized by a large DAOPHOT estimated error) in comparison 
with the number of good points, and to the
significant aliases produced by the data sampling. The bad points are
produced essentially by two causes: crowding, which implies a bad 
identification of the stars related to the variable seeing conditions, 
and occasional slight deformations of one stellar image which is interpreted 
by DAOPHOT as two close stars. The average number of bad points which must be 
discarded for obtaining a reasonable time series is about 2 per star in the
case of Cepheids, and the maximum number of discarded points is 6.
In general, stars with not less than 24 data points have been considered.

The data sampling is such that often there are significant aliases.
Fig. 6 shows the spectral window. As we can see there is a strong alias
with a complex structure at 1 c/d, so that if $P=1/f$ is 
the true period, we should expect strong aliases at the frequencies 
$1+f$ and $1-f$. Usually there are not significant problems for intermediate 
periods. Often a visual comparison of the data phased with the different 
altenatives is sufficient to solve the possible ambiguities.
On the other hand the aliasing makes it difficult to discriminate 
between long periods (say $P>60$ d) and  periods very close 
to 1 d  ($0.983<P<1.017$ d), even if it appears more reasonable to expect,
at least on a statistical basis, that most of these stars are long
period objects. Another relevant ambiguity regards the possible very short period
Cepheids ($P<1$ d), for which, due the generally small amplitudes
and therefore the low $S/N$, it is difficult to judge if the best 
phasing is for $f>1$ c/d or $f-1$. 

\begin{figure}
\epsfxsize=12truecm
\epsffile[1 350 550 690]{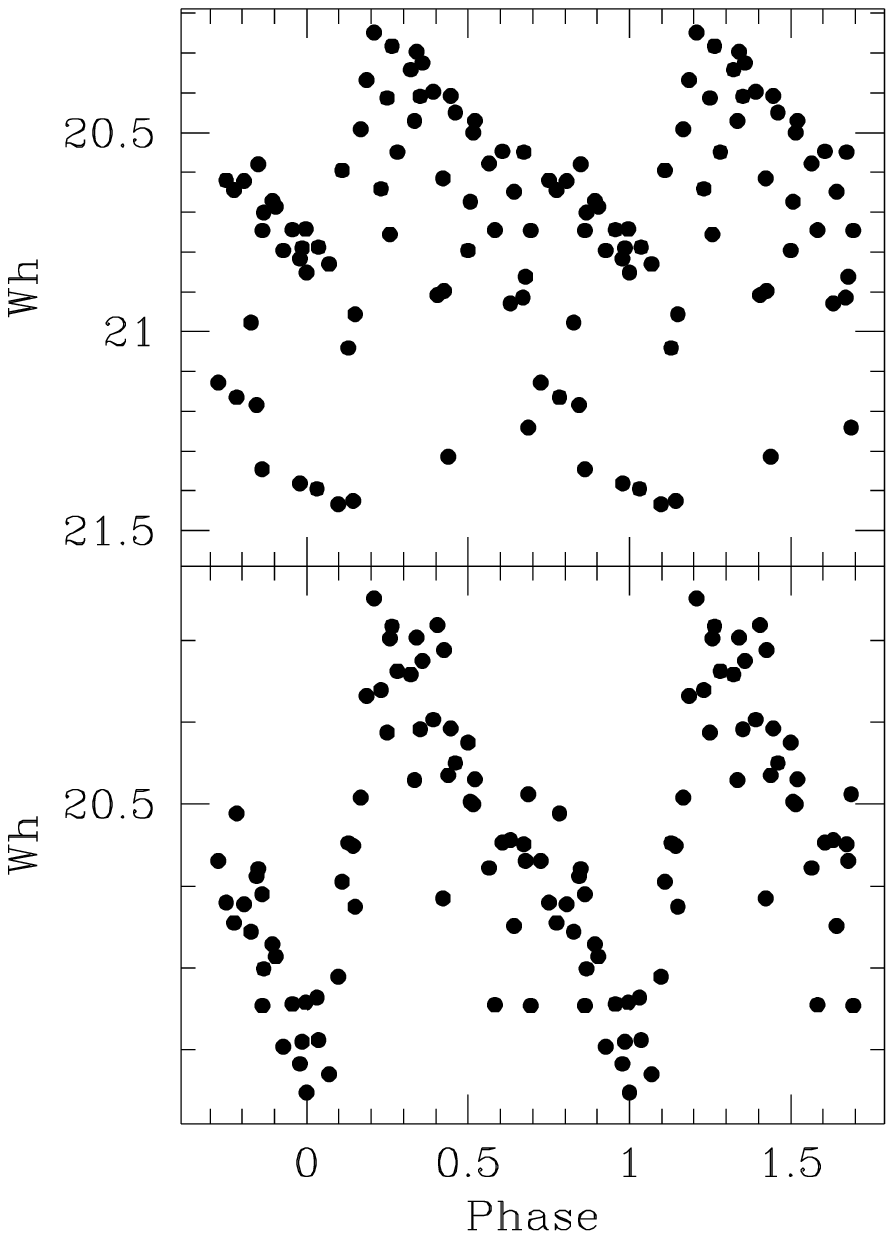}
\caption[ ]{$Wh$ light curve of the Cepheid V2221 showing the effect
of a close star (about 0{\farcs}8). Upper panel: in many cases the
two stars are not discriminated by DAOPHOT; lower panel: when resolved,
the luminosities of the two stars are summed in order to get a consistent 
light curve
}  
\end{figure}

\begin{figure}
\epsfxsize=7truecm
\epsffile[30 170 450 690]{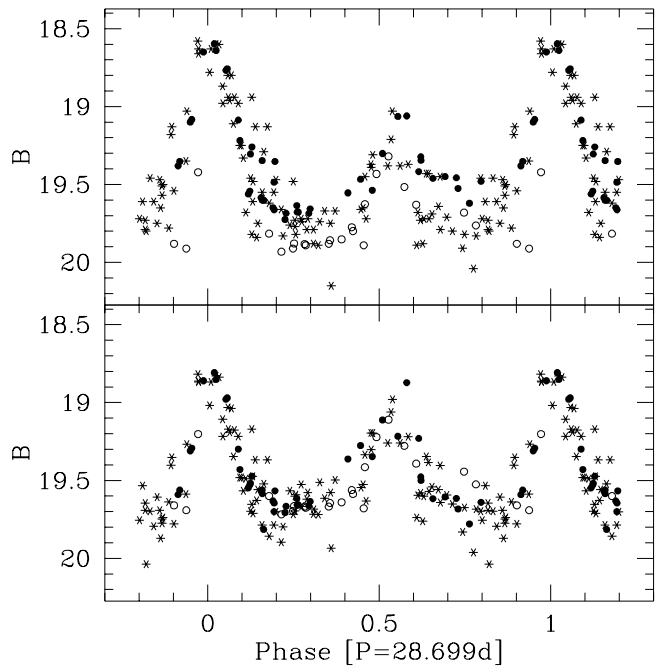}
\caption[ ]{Light curve of V1740 (SV39). {\em Filled circles:} 1995--1997
$Wh$ data; {\em open circles:} 1998 $Wh$ data; {\em asteriscs:} photographic 
$B$ data published by Sandage (1971); the $Wh$ data have been rescaled
(see text). Upper panel: data phased with $P=28.699$ d; lower panel: 
data phased with the same period after subtracting the component with
$P=1123$ d
}  
\end{figure}

\begin{table}
\caption[]{Variable stars in common with Baade-Sandage}
\begin{flushleft}
\begin{tabular}{lllll}
\hline\noalign{\smallskip}
Baade--  & period [d]  & type & present & period or  \\
Sandage &             &      &  work    & timescale [d]  \\
\noalign{\smallskip}\hline\noalign{\smallskip}
  SV9  & 5.58   & Cep  & V0819  & 5.578 \\
 SV10  & 4.065  & Cep  & V1897  & 4.065 \\
 SV17  & 5.73   & Cep  & V1734  & 5.737 \\
 SV18  & 16.43  & Cep  & V1039  & 16.43  \\
 SV21  &        & Irr  & V3106  & $>1000$ \\
 SV22  & 146.3  & Cep  & V2396  & (145.6) \\ 
 SV24  & 6.74   & Cep  & V1337  &  6.743 \\
 SV29  & 2.869  & Cep  & V0555  &  2.868 \\
 SV32  &        & Irr  & V0089  & $\sim650$ \\
 SV36  & 2.444  & Cep  & V1756  & 2.445 \\
 SV38  &        & Irr    & V0076  & $>1000$  \\
 SV39  & 28.72  & (Cef?) & V1740  &  *      \\
 SV40  &        & Irr    & V1872  & $\sim650$ \\
 SV43  &        & Irr  & V2321  & $\sim 62$ \\
 SV52  &        & Irr  & V1908  & $\sim$300? \\
 SV54  & 3.663  & Cep  & V2256  & 3.663 \\
 SV56  &        & ?    & V1800  &  *    \\    
\noalign{\smallskip}
\hline
*See text
\end{tabular}
\end{flushleft}
\end{table}

\section{Results}

About 110 stars were detected as variables: population I and II Cepheids, 
eclipsing binaries, long period, semiregular and irregular variables. 
In Table 3 we have reported the stars discovered by Baade and by Sandage 
(Sandage \cite{san}) falling in Field A, their period, variable type, the 
identification number in the present work and the period obtained by us. 
All the previously known Cepheids are confirmed; the periods obtained with our 
data are very similar to the old ones. No attempt has been done for increasing 
the significant digits of the periods by analysing old and new data together. 
For the irregular variables we have reported the time scale of variability 
which is compatible with our data. Some of the stars are discussed in 
the following subsections; in particular, the unusual characteristics of the star 
SV39 (V1740) are also confirmed, and they are discussed in Sect. 5.4.

Cepheids, other periodic variables, eclipsing binaries and irregular or
semiregular variables are listed in Table 4, 5, 6 and 7, respectively.
The stars are identified by their name; the right ascension and
declination are given along with the $P$ and the mean $Wh$ magnitude
(for eclipsing binaries an estimate of $Wh$ at the maximum luminosity is
reported). For Cepheids the probable pulsation mode is also reported, while
for the irregular, possible long period and semiregular variables
(Table 7) the timescale of variability is indicated. 
The astrometric positions were computed using 11 previously known variable stars 
as local astrometric standards, to derive transformation equations from 
the CCD $(x,y)$ positions to $\alpha$(1950) and $\delta$(1950). 
The coordinatae were taken from the General 
Catalogue of Variable Stars (GCVS; Samus \cite{gcvs}). Other four stars,
SV36 (V1756), SV40 (V1872), SV52 (V1908) and SV43 (V2321), were excluded 
because the $\delta$ values in GCVS differ by some arcsecs from 
the results of the transformation equations; this indicates an 
identification problem. The comparison of the derived positions with those 
reported by Freedman (\cite{fre2}) shows that the accuracy of the 
transformation is generally better than about 0{\farcs}5.

The variable stars with known $V-R$ index are shown in the color-magnitude
diagram of Fig. 7.

\subsection{Cepheids}
The light curves of the detected Cepheids are displayed in Fig. 8 and 9;
note that the magnitude scale is not the same in the different panels. 
Just from a simple inspection it is possible to conclude that both 
fundamental and first overtone mode Cepheids have been detected;
the fundamental mode Cepheids have large amplitude or asymmetric light
curves, while first overtone modes have relatively small amplitude and more 
symmetric light curves. We have used the Fourier parameters and the amplitudes
for discriminating the pulsation mode. A detailed discussion of these parameters 
and comparisons with other galaxies will be reported in Paper II 
(Antonello et al \cite{a2}). There is no reliable indication of 
double-mode Cepheids; probably the precision and the sampling of the data are 
not sufficient for their detection.

The stars with known $V$--$R$ occupy a vertical band in the color-magnitude
diagram (Fig. 7), or instability strip. The $PL$ diagram for the $Wh$-band is 
briefly discussed in Sect. 6.2 and shown in Fig. 15. 
\begin{table}
\caption[]{Cepheids in Field A of IC1613}
\begin{flushleft}
\begin{tabular}{llllll}
\hline\noalign{\smallskip}
Name & $\alpha$(1950) & $\delta$(1950) & P & $<$Wh$>$ & Puls.  \\
     &  [$^h$~~$^m$~~$^s$]       & [$^o$~~ '~~ '']  &  [d]  &  & mode \\
\noalign{\smallskip}\hline\noalign{\smallskip}
 V0078 & 1 2 26.3 & +1 52 30.5 &     1.435 & 22.15 &F \\
 V0107 & 1 2 22.6 & +1 52 33.2 &     6.714 & 20.25 &F \\
 V0128 & 1 2 25.9 & +1 52 53.4 &     1.251 & 22.34 &1-O \\
 V0178 & 1 2 22.4 & +1 52 53.8 &     0.817 & 22.40 &... \\
 V0236 & 1 2 24.5 & +1 53 19.2 &     1.390 & 22.26 &... \\
 V0279 & 1 2 30.5 & +1 53 41.5 &     2.098 & 22.46 &F \\
 V0368 & 1 2 27.2 & +1 53 58.1 &     1.479 & 22.05 &1-O \\
 V0377 & 1 2 20.6 & +1 53 57.3 &     1.643 & 22.23 &... \\
 V0414 & 1 2 16.8 & +1 54 00.5 &     2.459 & 22.27 &F \\
 V0479 & 1 2 27.7 & +1 54 28.8 &     0.663 & 22.35 &... \\
 V0524 & 1 2 17.4 & +1 54 36.6 &     1.424 & 22.49 &... \\
 V0533 & 1 2 30.6 & +1 54 49.9 &     1.845 & 22.45 &F \\
 V0551 & 1 2 26.4 & +1 54 59.1 &     2.690 & 22.28 &... \\
 V0555 & 1 2 30.9 & +1 55 00.0 &     2.868 & 21.49 &F \\
 V0655 & 1 2 21.7 & +1 55 23.1 &     2.538 & 22.41 &... \\
 V0819 & 1 2 21.9 & +1 52 13.7 &     5.578 & 21.06 &F \\
 V1014 & 1 2 29.3 & +1 54 18.2 &     2.950 & 21.97 &F \\
 V1039 & 1 2 28.5 & +1 54 27.8 &    16.431 & 19.82 &F \\
 V1100 & 1 2 26.6 & +1 55 06.6 &     1.287 & 22.16 &1-O \\
 V1178 & 1 2 22.3 & +1 55 50.1 &     1.628 & 22.36 &F \\
 V1241 & 1 2 27.0 & +1 53 07.8 &     1.049 & 21.95 &1-O \\
 V1289 & 1 2 25.3 & +1 54 00.9 &     0.646 & 22.33 &... \\
 V1296 & 1 2 17.6 & +1 54 03.2 &     0.859 & 22.53 &... \\
 V1337 & 1 2 17.4 & +1 54 47.1 &     6.743 & 20.45 &F \\
 V1371 & 1 2 26.8 & +1 55 31.3 &     0.886 & 22.38 &1-O \\
 V1479 & 1 2 28.3 & +1 52 48.7 &     1.123 & 22.15 &1-O \\
 V1592 & 1 2 26.7 & +1 53 04.4 &     4.360 & 21.76 &F \\
 V1734 & 1 2 17.3 & +1 54 02.4 &     5.737 & 20.68 &F \\
 V1756 & 1 2 18.9 & +1 54 14.6 &     2.445 & 22.27 &F \\
 V1767 & 1 2 20.2 & +1 54 21.8 &     0.797 & 22.31 &2-O? \\
 V1798 & 1 2 26.3 & +1 54 47.1 &     3.272 & 21.68 &F \\
 V1897 & 1 2 25.5 & +1 52 17.2 &     4.065 & 20.92 &F \\
 V2020 & 1 2 29.0 & +1 54 43.5 &     1.870 & 22.33 &1-O \\
 V2100 & 1 2 30.1 & +1 52 53.9 &     1.879 & 22.69 &... \\
 V2150 & 1 2 18.7 & +1 53 48.9 &     2.567 & 21.47 &F \\
 V2172 & 1 2 23.5 & +1 54 31.8 &     1.431 & 22.12 &1-O \\
 V2221 & 1 2 30.3 & +1 52 43.4 &     5.721 & 21.12 &F \\
 V2256 & 1 2 28.0 & +1 53 23.2 &     3.663 & 21.29 &F \\
 V2262 & 1 2 27.5 & +1 53 30.1 &     1.443 & 21.71 &... \\
 V2309 & 1 2 19.0 & +1 54 55.6 &     2.180 & 21.95 &F \\
 V2396 & 1 2 26.4 & +1 54 40.7 &   145.6   & 17.43 &F \\
 V2414 & 1 2 25.4 & +1 52 35.5 &     7.573 & 20.83 &F \\
 V2766 & 1 2 21.7 & +1 52 09.9 &     2.238 & 21.84 &F \\
\noalign{\smallskip}
\hline
\end{tabular}
\end{flushleft}
\end{table}

In the following we report some notes on selected stars.

{\em V2396}. The data sampling does not allow to construct the 
complete light curve, however the period found by us is close to 
that given by Sandage (\cite{san}).

{\em V2414} and {\em V0107} have very small amplitudes, about 0.2 - 0.3
mag; looking at our Galaxy, this is not unusual for stars with $P$ between 
7 and 10 d.

{\em V1337} has the best light curve, which can be fitted with a
5th order Fourier decomposition and rms residual of 0.026 mag.

{\em V2221} is an interesting case, because it has a close (0{\farcs}8) 
companion which is slightly fainter. DAOPHOT was not able to resolve always
the two stars, and therefore the resulting light curve was very scattered.
We have simply summed the luminosities of the two stars when they were 
resolved; the two light curves are shown in Fig. 10. Some scatter is still 
present, but the Cepheid behavior is evident; clearly its amplitude is smaller 
than what should be expected, and one should correct for the companion's 
luminosity before using the star in a $PL$ relation. 
The referee has remarked, however, that these problems given by close stars 
can be overcome by using the fixed position photometry 
(see Kaluzny et al. 1998).

{\em V1592} was not noticed by Baade and Sandage, even if it has large
amplitude and is sufficiently bright, probably because it is located
in a partially crowded region.

{\em V0551, V0655} and {\em V2100} are characterized by rather symmetric
light curves, but their periods and luminosities are typical of 
fundamental mode Cepheids; the available color of one of them indicates 
a location in the instability strip. The nature of these stars is
uncertain; tentatively we put them in relation with the anomalous Cepheids,
even if their periods and luminosities are larger than those seen in
galactic and extragalctic anomalous Cepheids.

\begin{figure*}
\epsfxsize=14truecm
\epsffile[30 180 450 690]{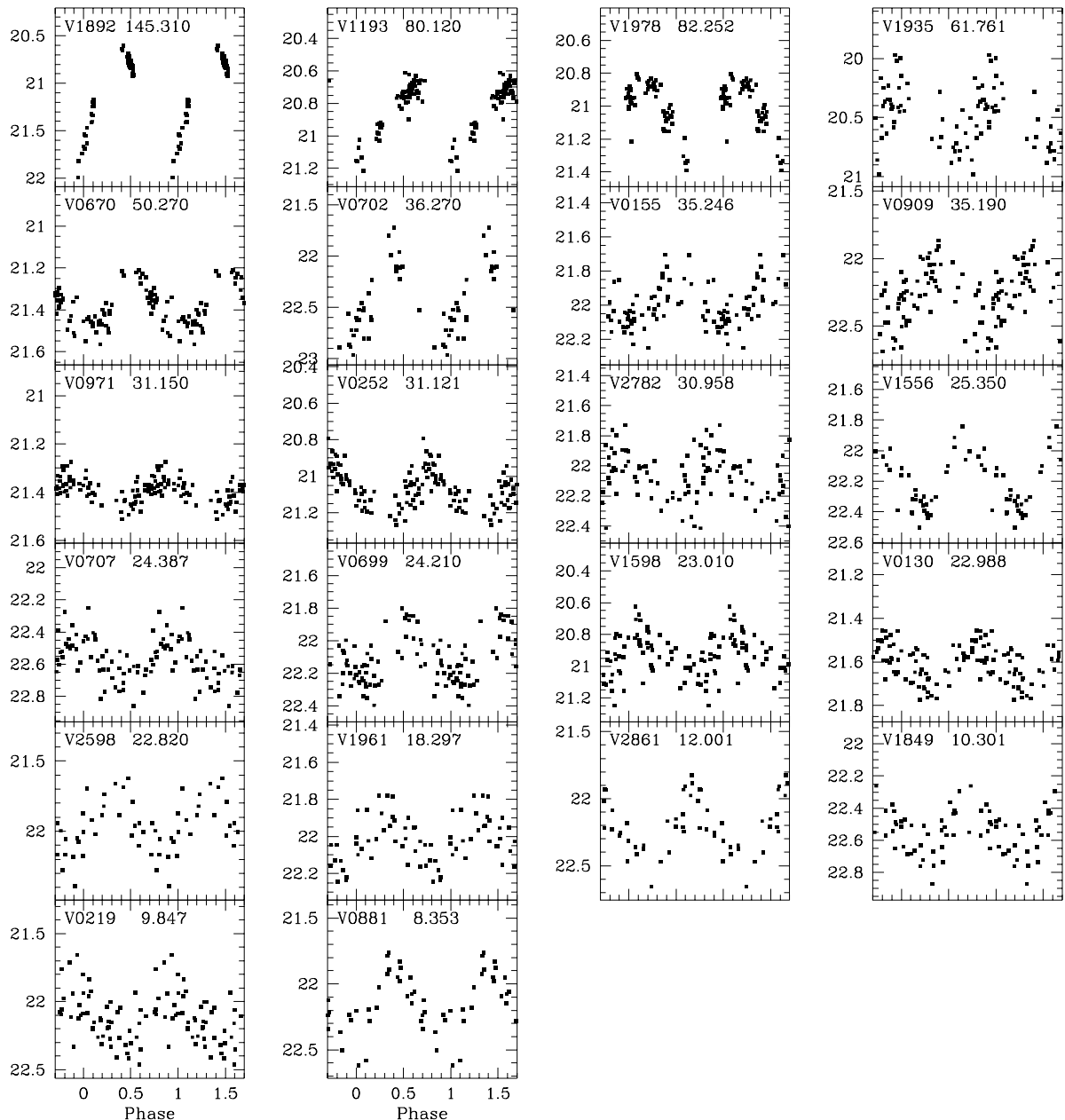}
\caption[ ]{$Wh$ light curves of periodic variables.
}  
\end{figure*}

\begin{figure}
\epsfxsize=14truecm
\epsffile[30 200 450 690]{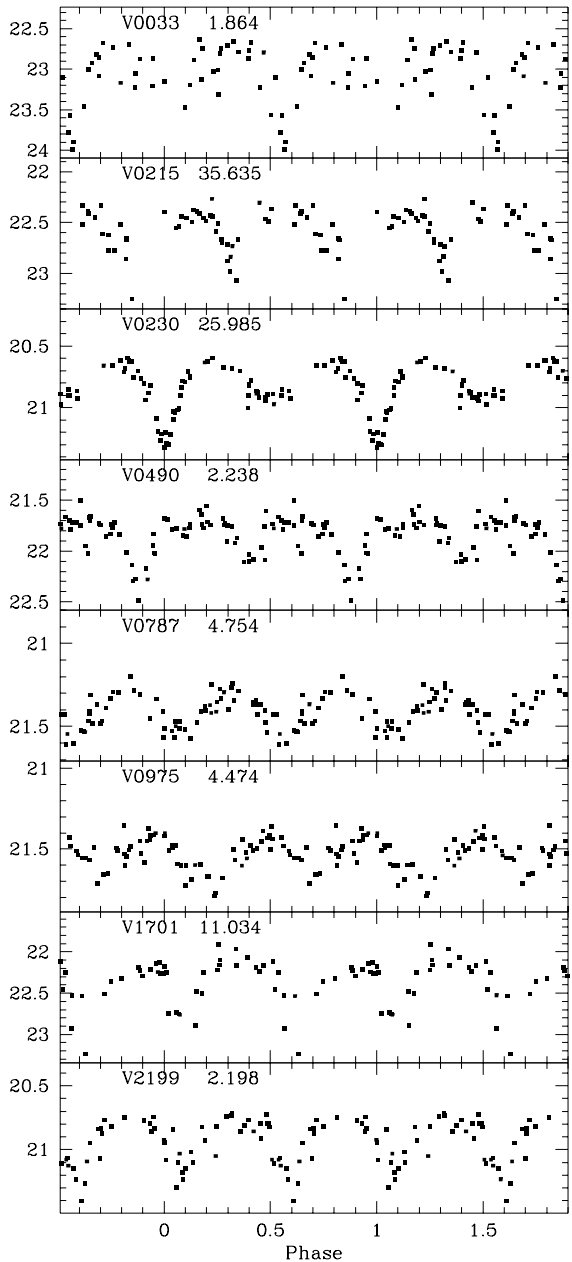}
\caption[ ]{$Wh$ light curves of eclipsing binaries.
}  
\end{figure}

\subsection{SV39=V1740}

The enigmatic nature of this variable was already pointed out by 
Sandage (1971). According to the old data the light curve could be described 
as an inverted $\beta$ Lyrae eclipsing variable with a period of 28.72 d. 
Notwithstanding this very peculiar shape and the high luminosity with respect 
to the $PL$ relation, Sandage included it among Cepheids and as such it 
remains in the GCVS. We have reanalyzed by 
means of the power spectrum technique both the old photograpic measurementes 
(107 datapoints) and our 67 $Wh$ ones. The analysis was performed both 
separately for the two data sets and by merging the sets after rescaling the 
$Wh$ data to the $B$ ones (the scale factor is 1.9) and aligning the 
zeropoints of the two timeseries. The light variation seems rather complicated
but a firm conclusion can be reached: there are at least two periodic terms 
which fit both datasets, one with a period of 28.699 d, very similar to that 
suggested by Sandage, and the other with P=1123 d. The reality of this long 
period term can be deduced from Fig. 11. In the upper panel we have plotted
all the data phased with the short period. The photographic data are indicated
by asteriscs, while filled circles represents the $Wh$ data of 
the season 1995--97, and open circles those of 1998. It is evident that in 
1998 the star was sistematically less bright than in 1995--97. In the lower panel we 
have subtracted from the data the long period term. Now the data dispersion about 
a mean curve of both photographic and $Wh$ data is considerably decreased, 
and in particular the 1998 $Wh$ data are well aligned with those of previous 
years. Another fact can be deduced from the figure: after removing the long 
period term, the difference between the two maxima is significantly decreased.
As a matter of fact if we adopt as the short period 14.350 d instead of its
double, the fit of the data is only marginally worse. Therefore with the 
present available dataset, we prefer to be cautios and not do decide 
which of the two possible short periods is the correct one.
Another and more pregnant open question is the nature of the variability.
The color of the star, $V$--$R$=0.52, is similar to that of Cepheids;  however 
no radially pulsating star with such an amplitude is known to have 
symmetric maxima, and furthermore it is not possible to explain theoretically 
such a shape. A check of the amplitudes is not much conclusive: assuming the amplitude
was essentially constant during fifty years, the ratio of photographic and 
white light amplitudes is in the range 0.9 - 1.5 for 5 bright Cepheids,
and only V1039 has ratio 1.9 as V1740.

\begin{table}
\caption[]{Other periodic variables in Field A of IC1613}
\begin{flushleft}
\begin{tabular}{llllll}
\hline\noalign{\smallskip}
Name & $\alpha$(1950) & $\delta$(1950) & P & $<$Wh$>$ & Type \\
     &  [$^h$~~$^m$~~$^s$]       & [$^o$~~ '~~ '']  &  [d]  &        \\
\noalign{\smallskip}\hline\noalign{\smallskip}
 V0130 & 1 2 23.8 & +1 52 42.8 &    22.99 & 21.61 & W Vir \\ 
 V0155 & 1 2 24.4 & +1 52 51.0 &    35.27 & 22.00 &\\
 V0219 & 1 2 22.6 & +1 53 11.6 &     9.847 & 22.11 &\\
 V0252 & 1 2 28.4 & +1 53 26.5 &    31.12 & 21.06 &\\
 V0670 & 1 2 17.6 & +1 55 29.1 &    50.2 & 21.38 &\\
 V0699 & 1 2 20.9 & +1 55 34.9 &    24.2 & 22.14 &\\
 V0702 & 1 2 25.9 & +1 55 39.3 &    36.27 & 22.56 &\\
 V0707 & 1 2 27.7 & +1 55 42.3 &    24.4 & 22.59 &\\
 V0881 & 1 2 27.8 & +1 53 03.4 &     8.353 & 22.12  & W Vir\\ 
 V0909 & 1 2 30.1 & +1 53 16.1 &    35.18 & 22.27 &\\
 V0971 & 1 2 24.9 & +1 53 51.4 &    31.15 & 21.39  &  W Vir\\  
 V1193 & 1 2 18.4 & +1 52 10.5 &    80.1 & 20.81  & LP, RV \\
 V1556 & 1 2 19.2 & +1 52 47.6 &    25.35 & 22.22 &\\
 V1598 & 1 2 26.0 & +1 53 06.1 &    23.01 & 20.94  &  W Vir\\
 V1849 & 1 2 21.6 & +1 55 24.4 &    10.30 & 22.54 &\\
 V1892 & 1 2 22.3 & +1 52 09.3 &   145.1 & 21.03  & LP \\
 V1935 & 1 2 24.1 & +1 52 59.0 &    61.66 & 20.54  &  W Vir\\
 V1961 & 1 2 18.7 & +1 53 23.5 &    18.4 & 22.01 &\\
 V1978 & 1 2 25.3 & +1 53 50.5 &    82.0 & 20.99  & LP \\
 V2598 & 1 2 27.2 & +1 53 09.6 &    22.82 & 21.97 &\\
 V2782 & 1 2 29.3 & +1 52 25.4 &    30.96 & 22.05 &\\
 V2861 & 1 2 16.7 & +1 52 49.0 &    12.0 & 22.19 &\\
\noalign{\smallskip}
\hline
\end{tabular}
\end{flushleft}
\end{table}

\subsection{Other periodic variables}
Periodic variables include red variables and population II pulsating stars 
(Fig. 12), and eclipsing binaries (Fig. 13). 

{\em V1892, V1193} and {V1978} are probable long period variables.
In particular, V1193 could be an RV Tau star with a long period of
160.2 d; however $V$--$R$ is 0.94. 

The population II Cepheids (or W Vir stars) should have $V$--$R$ not 
very different from  that of population I Cepheids, since the latter 
are rather metal-poor;
moreover, for a given luminosity, population II Cepheids have a much longer 
period. On these basis we have identified 5 of such stars, namely
{\em V00130, V0881, V0971, V1598} and {\em V1935}.

\begin{table}
\caption[]{Eclipsing binaries in Field A of IC1613}
\begin{flushleft}
\begin{tabular}{lllll}
\hline\noalign{\smallskip}
Name & $\alpha$(1950) & $\delta$(1950) & P & $Wh_{max}$  \\
     &  [$^h$~~$^m$~~$^s$]  & [$^o$~~ '~~ '']  &  [d]  &        \\
\noalign{\smallskip}\hline\noalign{\smallskip}
 V0033 & 1 2 26.5 & +1 52 18.6 &     1.864 & 22.7 \\
 V0215 & 1 2 16.5 & +1 53 06.2 &    35.635 & 22.4 \\
 V0230 & 1 2 25.6 & +1 53 27.1 &    25.985 & 20.6 \\
 V0490 & 1 2 29.6 & +1 54 33.8 &     2.238 & 21.7 \\
 V0787 & 1 2 24.1 & +1 52 17.2 &     4.754 & 21.3 \\
 V0975 & 1 2 19.9 & +1 53 46.1 &     4.474 & 21.4 \\
 V1701 & 1 2 30.7 & +1 53 52.2 &    11.034 & 22.1 \\
 V2199 & 1 2 29.1 & +1 52 13.2 &     2.198 & 20.8 \\
\noalign{\smallskip}
\hline
\end{tabular}
\end{flushleft}
\end{table}

From a simple inspection of the phased light curves we have identified
24 possible eclipsing binaries. Only 8 stars are reported in Table 6
and are shown in Fig. 13. A detailed anlaysis of all the candidates
will be performed in a subsequent paper.

\subsection{Other semiregular and irregular variables}

There are several stars which are characterized by irregular
variability on different time scales; some of them could be long
period variables which cannot be identified as such owing to the
data sampling and the short observing time interval. In Fig. 14 some
of these stars are shown.

{\em SV21=V3106} could vary both with short (tens of days) and long 
($\sim 1000$ d) timescales; the amplitude is about 1 mag and the color is
$V$--$R \sim 0.1$.

{\em SV52=V1908} is variable with small amplitude (less than 0.3 mag), 
and appears brighter than fifty years ago. The color $V$--$R$=0.39 
indicates a yellow star in the upper part of the instability strip.

If the identification is correct, the star {\em SV56=V1800} previously
known as irregular variable, appears to be constant or variable with very 
small amplitude (less than about 0.1 mag); it is not reported in Table 7.

\begin{table}
\caption[]{Irregular and possible long period and semiregular variables 
in Field A of IC1613}
\begin{flushleft}
\begin{tabular}{lllll}
\hline\noalign{\smallskip}
Name & $\alpha$(1950) & $\delta$(1950) &  $<Wh>$ & timescale of \\
     &  [$^h$~~$^m$~~$^s$]       & [$^o$~~ '~~ '']  &   & variability [d]\\
\noalign{\smallskip}\hline\noalign{\smallskip}
 V0076 & 1 2 27.3 & +1 52 31.6 & 16.32 & 1000\\
 V0089 & 1 2 24.1 & +1 53 00.7 & 16.73 &  650\\
 V0106 & 1 2 29.2 & +1 52 41.5 & 22.12 &  150\\
 V0157 & 1 2 23.8 & +1 52 48.6 & 21.63 &  220\\
 V0193 & 1 2 16.5 & +1 52 49.9 & 20.89 &  260\\
 V0218 & 1 2 18.6 & +1 53 07.9 & 22.19 &  280\\
 V0253 & 1 2 17.9 & +1 53 17.1 & 21.25 &   62\\
 V0496 & 1 2 16.4 & +1 54 34.7 & 21.49 &   42\\
 V0505 & 1 2 23.6 & +1 54 36.7 & 21.00 &  130\\
 V0530 & 1 2 20.2 & +1 54 43.0 & 20.78 &   54\\
 V0663 & 1 2 26.9 & +1 55 35.3 & 21.70 &  310\\
 V0688 & 1 2 23.4 & +1 55 34.5 & 22.26 &  660\\
 V0764 & 1 2 16.5 & +1 52 01.0 & 21.98 &  110\\
 V0841 & 1 2 16.4 & +1 52 29.1 & 21.61 &  120\\
 V0842 & 1 2 31.4 & +1 52 39.7 & 22.28 &   46\\
 V0861 & 1 2 26.6 & +1 52 47.9 & 21.53 &   80\\
 V0929 & 1 2 18.9 & +1 53 16.5 & 21.20 &   77\\
 V1130 & 1 2 28.7 & +1 55 27.5 & 21.32 &   18\\
 V1167 & 1 2 19.4 & +1 55 39.6 & 19.33 &   56\\
 V1177 & 1 2 21.5 & +1 55 48.9 & 21.04 &     \\
 V1181 & 1 2 22.7 & +1 55 53.8 & 20.82 &  990\\
 V1224 & 1 2 18.1 & +1 52 46.0 & 22.17 &  550\\
 V1322 & 1 2 19.1 & +1 54 35.5 & 21.04 &  480\\
 V1331 & 1 2 24.6 & +1 54 46.6 & 20.71 &  100\\
 V1758 & 1 2 17.3 & +1 54 15.3 & 21.18 &   46\\
 V1783 & 1 2 23.2 & +1 54 32.5 & 21.44 &  430\\
 V1812 & 1 2 21.0 & +1 54 51.9 & 20.86 &  630\\
 V1830 & 1 2 21.2 & +1 55 06.5 & 21.23 &  104\\
 V1872 & 1 2 29.5 & +1 55 49.3 & 19.37 &  650\\
 V1908 & 1 2 18.9 & +1 52 21.4 & 20.14 &  300\\
 V2033 & 1 2 26.5 & +1 55 05.5 & 20.75 &  100\\
 V2095 & 1 2 25.0 & +1 52 46.6 & 21.26 &  270\\
 V2109 & 1 2 21.6 & +1 53 00.9 & 21.80 &  230\\
 V2124 & 1 2 26.0 & +1 53 21.6 & 20.56 &  440\\
 V2178 & 1 2 17.2 & +1 54 33.6 & 20.73 &  120\\
 V2321 & 1 2 30.8 & +1 55 47.4 & 17.30 &   62\\
 V2521 & 1 2 28.6 & +1 52 44.5 & 21.78 &   12\\
 V3106 & 1 2 30.7 & +1 54 44.3 & 20.31 & 1000\\
\noalign{\smallskip}
\hline
\end{tabular}
\end{flushleft}
\end{table}

\begin{figure*}
\epsfxsize=14truecm
\epsffile[30 180 450 690]{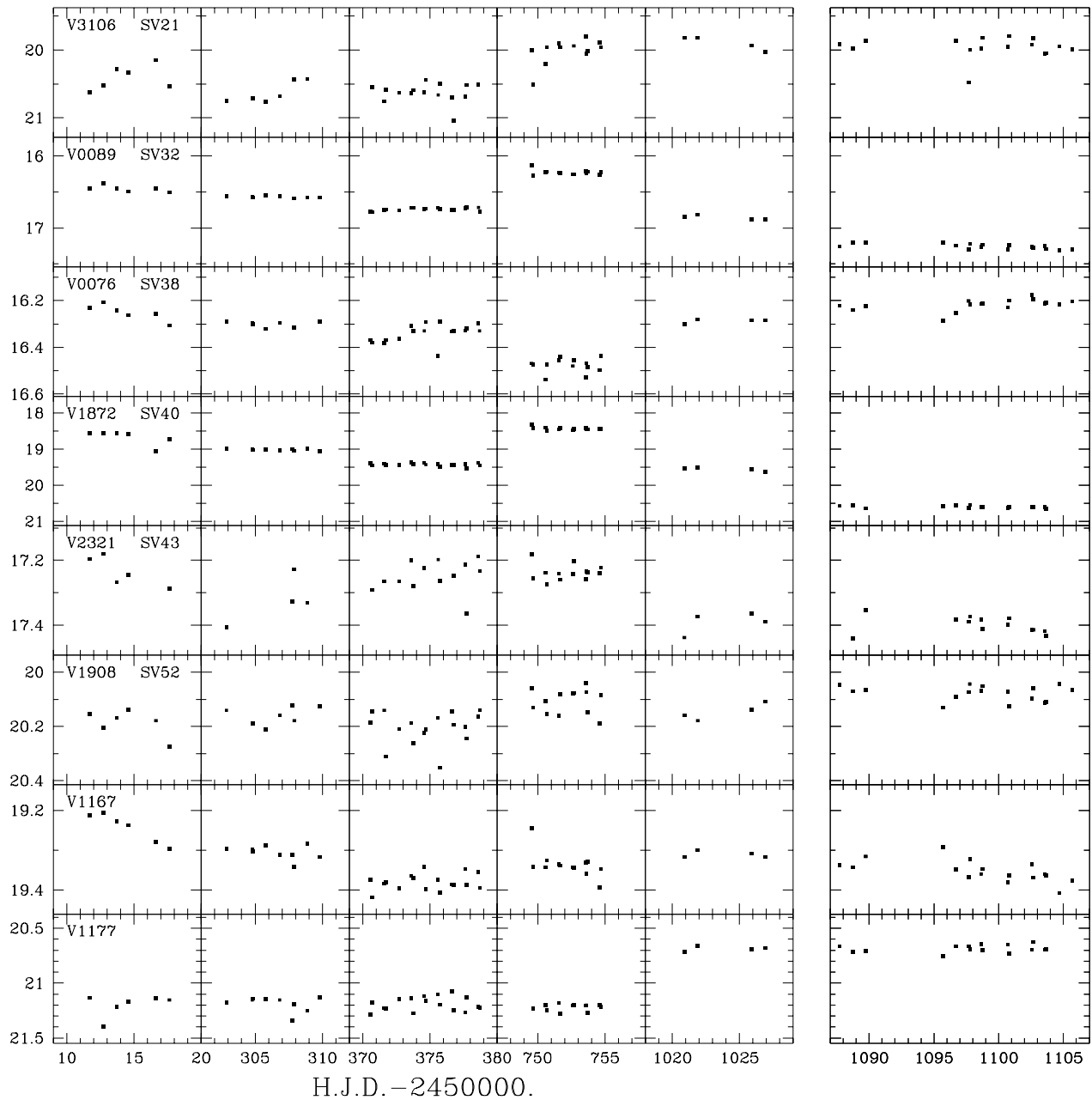}
\caption[ ]{$Wh$ light curves of a sample of irregular variables;
note the different magnitude scales
}  
\end{figure*}

\section {Discussion}

\subsection{The $Wh$-band}
The main aim of our research is the study of Cepheid light curves 
in nearby galaxies. This study requires a certain amount of observing telescope
time, and owing to the pressure on the available observational facilities in
sites with good sky and seeing conditions, we were forced to exploit as much
as possible the relatively small telescopes. This reason, coupled with the
need of the highest signal-to-noise ratio for obtaining accurate light
curves, has implied our decision of observing with no filter. The advantage
with respect to usual filters in the optical range, such as Johnson or Gunn systems, 
is that the number of collected photons is larger by about a factor from 4 to 6,
which means that, for the same exposure time, an observation in $Wh$ band
with a 0.9 m telescope is equivalent, in terms of collected photons,
to an observation in $V$ band with a 2.1 m. On the other hand, the 
background sky tends to increase in the near infrared and this effect should be
more evident when observing galaxies with a strong red background. 
This is not the case of IC 1613, however, because we have estimated a
similar star/sky intensity ratio in $V$ and $Wh$ bands.

The observational data of IC 1613 Cepheids show that, for exposure 
times of half a hour with a 0.9 m telescope and a back-illuminated CCD, it is 
possible to get light curves which are as accurate as 0.03 mag for stars with 
$m_V \sim 21$ mag, and to detect variable stars as faint as $V \sim$ 23.

The photometric properties of the $Wh$ band appear reasonably good.
The effective wavelength for A-G spectral types is intermediate between that 
of Johnson $V$ and $R$ bands, and $Wh$ measurements correlate well with 
$V$ and $V$--$R$. The obvious defect is that the photometry depends on the 
instrument. For example the response of the system changes when
using a front-illuminated CCD instead of a back-illuminated one, because
the effective wavelength in the front-illuminated case is closer to that of 
$R$ band. Therefore some care will be required when merging {\em differential} 
$Wh$ observations of the same stellar field obtained with different instruments. 
Systematic effects related to star colours are expected, but they can be probably 
corrected for; this requires, however, one very deep exposure in $V$ and $R$ or
one observation with a larger telescope to get the colors of the faintest
stars. Star colors are important in any case for discussing the nature of
variables, therefore the suggested strategy for future work in this
field shall include at least one observation in a photometric system with
an adequately large telescope for obtaining this information.

It is interesting to compare, at least qualitatively, our results with those 
of analogous surveys such as DIRECT (Kaluzny et al. \cite{kal}), which is the 
project dedicated to the observations of M31. We recall that the distance of IC 1613,
$m-M=24.42$, is very similar to that of M31, $m-M=24.44$ (Madore \&
Freedman \cite{mf}), but the stars of M31 suffer of local reddening 
$E(B-V)$ from 0 to 0.25. DIRECT uses telescopes of 1.2 - 1.3 m,
front- and back-illuminated CCD detectors and exposure time of 900 s, and the 
number of collected photons would be approximately similar to that obtained 
by us with the 0.9 m telescope, exposure of 1800 sec and same filter. 
For a given period, the M31 Cepheid $V$ light curves appear less accurate 
than those of IC 1613 in $Wh$ band. Another indication is the faintest 
Cepheid with short period: in M31 the limit is about 4 d and $\sim 22$ mag 
(Stanek et al. \cite{sta}); in IC 1613 we {\em estimate} $V \sim 23$. 
We note that it is not easy to compare observations obtained with different telescopes, 
and the difference between M31 and IC 1613 results could depend in part also on 
crowding problems in M31, average seeing conditions, etc. In principle,
however, we expect a gain of about 1.5 mag due to the use of the $Wh$ band, for 
the same telescope and conditions.

\begin{figure}
\epsfxsize=21truecm
\epsffile[60 450 600 700]{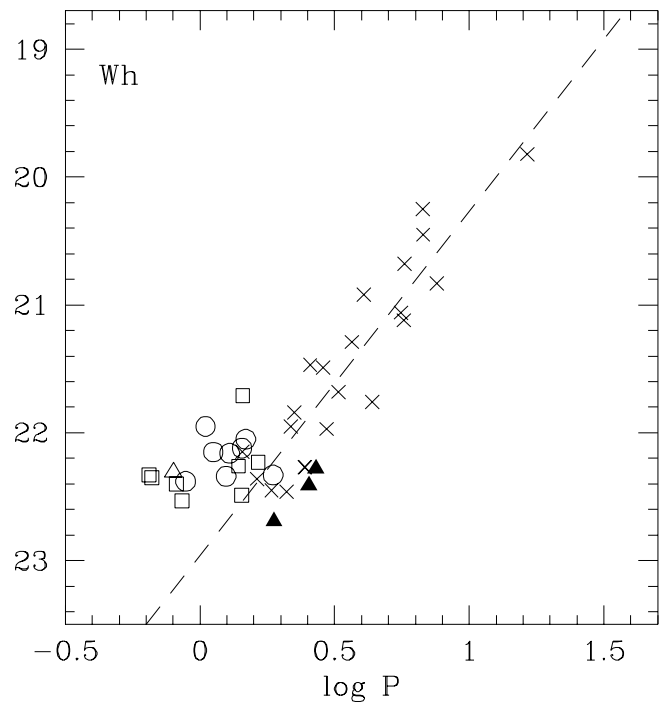}
\caption[ ]{$PL$ diagram for Cepheids in Field A of IC 1613.
{\em Crosses:} fundamental mode Cepheids; {\em open circles:} first overtone
mode Cepheids. The discrimination between the two modes was made on the
basis of the Fourier parameters and amplitudes. {\em Open squares:} Cepheids
with uncertain pulsation mode; most of them should be probable first overtone
mode pulsators. {\em Filled triangles:} stars with rather symmetric 
light curve and relatively long period (see Sect. 5.1). {\em Open triangle:} 
second overtone mode candidate. The dashed line is the statistical relation 
obtained for fundamental mode Cepheids
}
\end{figure}

\subsection{PL relation}

It is possible to derive a $PL$ relation for Cepheids using $Wh$ measurements,
and we should expect a similar slope to that obtained for $V$ and $R$. 
The relation is shown in Fig. 15 for 22 fundamental mode Cepheids (dashed line). 
The slope is --2.69$\pm$0.26 and the zero-point is 22.96$\pm$0.16; these 
figures were obtained without considering V2396 (with $P=145$d; but this 
exclusion is unessential) and we have reported them just for illustrative 
purposes. A detailed discussion will be made when the observations of all 
the fields of IC 1613 will be reduced; in particular, we will look for the 
possible bending of the relation at very short period, a feature which was 
observed in the Small Magellanic Cloud (Bauer et al. \cite{bau}).
The zero-point of our relation is obviously instrumentation dependent, 
even if in principle one could use a transformation such as eq. 2.
However the results of our study could be applied to distance 
determinations in another way. When Cepheids have been identified and their 
periods have been determined, then it is sufficient one observation in $V$ band
for constructing a standard $PL$ relation for the galaxy. Freedman 
(\cite{fre1}) has discussed and applied in detail this method, that is the
'single-phase' $PL$ relation. In Paper II we will show an example of
this application.

\section{Conclusion}
We have presented the first results of a survey of nearby galaxies for
detecting and studying Cepheids, and we have proven the utility of
the white-light-band $Wh$ observations, i.e. no filter, for reaching 
this goal. About 43 population I Cepheids with estimated luminosity as 
faint as $m_V \sim 23$ have been detected in Field A of IC 1613, while 
only 9 of them were previously known; most of the new stars have short 
period or small amplitude. 
A detailed study of these and of population II Cepheids will be reported
in Paper II.

The main conclusion of the present paper is the proposal of a new strategy for 
studying Cepheids as distance indicators. Relatively small telescopes
and the $Wh$-band can be used for the survey of nearby galaxies looking 
accurately for Cepheids. Then just one $V$, $R$ and/or $I$ images taken
with a two or three times larger telescope are needed in order to obtain 
the photometric data useful for the construction of the $PL$ relation for 
those galaxies. Of course, the procedure can be extended to far galaxies 
observed with comparatively larger telescopes. We are performing 
simulations for testing the advantage of this method over the standard 
technique when applied to VLT and HST.

\acknowledgements{The authors wish to thank the referee, K.Z. Stanek, for 
the useful comments.}

\end{document}